\documentclass{aastex}
\usepackage{spr-astr-addons}
\usepackage{url}
\usepackage{natbib}

\newcommand{\HI}{H\,{\sc i}}

\newcommand{\kms}{~km\,s$^{-1}$}
\newcommand{\kkms}{km\,s$^{-1}$}

\newcommand{\vsys}{$v_{\rm sys}$}

\newcommand{\MHI}{$M_{\rm HI}$}

\newcommand{\Msun}{~M$_{\odot}$}

\RequirePackage{color}

\newcommand{\emaila}{jsaponara@iar.unlp.edu.ar}

\begin{document}

\title{New \HI\ observations of KK\,69.\\ Is KK\,69 a dwarf galaxy in transition?}

\shortauthors{J.~Saponara et al.}

\author{J. Saponara\altaffilmark{1,2}} \and \author{B. S. Koribalski\altaffilmark{3}}
 \and \author{N. N. Patra\altaffilmark{4}} \and \author{P. Benaglia\altaffilmark{1,2}}
 
\affil{$^1$\,Instituto Argentino de Radioastronom\'{\i}a, (CONICET; CICPBA), C.C. No. 5, 1894,Villa Elisa, Argentina}
\affil{$^2$\,Facultad de Ciencias Astron\'omicas y Geof\'{\i}sicas, Universidad Nacional de La Plata, Paseo del Bosque s/n, 1900 La Plata, Argentina}
\affil{$^3$\,CSIRO Astronomy \& Space Science, Australia Telescope National Facility, P.O. Box 76, Epping, NSW 1710, Australia}
\affil{$^4$\,National Centre for Radio Astrophysics, Tata Institute of Fundamental Research (NCRA-TIFR), Pune 411 007, India}    

\email{\emaila}

\begin{abstract}

We present new \HI\ data of the dwarf galaxy KK\,69, obtained with the Giant Metrewave Radio Telescope (GMRT) with a signal-to-noise ratio that almost double previous observations. We carried out a Gaussian spectral decomposition and stacking methods to identify the cold neutral medium (CNM) and the warm neutral medium (WNM) of the \HI\ gas. We found that 30\% of the total \HI\ gas, which corresponds to a mass of $\sim$10$^{7}$\Msun, is in the CNM phase. The distribution of the \HI\ in KK\,69 is not symmetric. Our GMRT \HI\ intensity map of KK\,69 overlaid onto a Hubble Space Telescope image reveals an offset of $\sim$4~kpc between the \HI\ high-density region and the stellar body, indicating it may be a dwarf transitional galaxy. The offset, along with the potential truncation of the \HI\ body, are evidence of interaction with the central group spiral galaxy NGC\,2683, indicating the \HI\ gas is being stripped from KK\,69. Additionally, we detected extended \HI\ emission of a dwarf galaxy member of the group as well as a possible new galaxy located near the north-eastern part of the NGC\,2683 \HI\ disk.

\end{abstract}


\section{Introduction}

 The dwarf transition galaxies share properties with dwarf irregular (dIrr) galaxies (such us the \HI\ content) as well as with dwarf spheroidal (dSph) galaxies (low luminosity and old stellar populations, see for instance \cite{2003AJ....125.1926G}). In most of these galaxies, a positional offset is observed between the stellar and the \HI~component. \cite{1998ARA&A..36..435M}, \cite{2003AJ....125..593S}, and \cite{2003AJ....125.1926G} proposed a classification scheme for dwarf transition galaxies based on \HI~and $\rm H_\alpha$ flux. According to this classification, dwarf transition galaxies are detected in \HI~with very little or no $\rm H_\alpha$ flux indicating a gas depletion timescale higher than 100 Gyr. Even though the origin of the dwarf transition galaxies are not fully understood, many studies contemplate the idea that these galaxies could be the precursors of the dSph galaxies \citep[see, e.g.][]{2012ApJS..198....2K,2009ApJ...696..385G}. Hence, the dwarf transition galaxies could serve as a link between the late-type and early-type dwarf galaxies.
 
The shallow potential well of dwarf galaxies makes their interstellar medium highly susceptible to disruptions by environmental forces as well as internal processes. The star formation bursts are the primary internal reason of disarrangement and mass loss in dwarf galaxies \citep{1974MNRAS.169..229L,1986ApJ...305..669V,1986ApJ...303...39D}, while the ram-pressure stripping and tidal interactions are the main external mechanisms of gas removal. The efficiency of ram-pressure stripping depends on the intragroup medium density and the velocity of the galaxy through the medium \citep{1972ApJ...176....1G}; it is common to observe a tail feature in the opposite direction to the galaxy motion \citep{2007ApJ...659L.115C}. The magnitude of tidal disruptions depend on the mass, relative velocity and orbits between the interacting galaxies. The stripped gas may be found as tidal tails, filaments and even bridges of stars and gas \citep{1996ApJ...467..241H,2003MNRAS.339.1203K,2004MNRAS.348.1255K,2010AJ....139..102E}. The galaxies Phoenix and NGC\,5237 are two examples of that. Both are dwarf transition galaxies where the \HI\ gas is offset from the stellar component, but the origin of the displacement seems to be different. In Phoenix, the offset may be the result of the star formation activity \citep{2007ApJ...659..331Y}. However in NGC\,5237, the environmental effects are suspected to be the dominant reason of the offset \citep{2018MNRAS.tmp..467K}.\\

The internal and external mechanisms mentioned are not only the cause of structural/morphological changes, but they may also influence the future star formation activity in galaxies. \cite{2018A&A...612A..26A} have found that almost 10$\%$ of the Leo\,T total mass is in the cold \HI\ phase; conversely, the galaxy is not forming stars intensively. The authors attributed the large cold \HI\ gas detected as a result of the interaction of Leo\,T with the Milky Way circumgalactic medium. The \HI\ gas constitutes a key component in the evolution of galaxies; especially the CNM phase which is often use to pin-point the future star formation regions. Thus, the presence of the CNM and the columnar density of it, among other parameters, is crucial to better understand the star formation activity in dwarf transition and dIrr galaxies.\\

The dwarf galaxy KK\,69 is part of the Faint Irregular Galaxies GMRT Survey \citep[FIGGS]{2008MNRAS.386.1667B} and is the only one with characteristics of a dwarf transition galaxy. The velocity resolution of all FIGGS galaxies is 1.7\kms. However, the signal-to-noise (SNR) achieved to some of them is not sufficient to perform the decomposition of the cold and warm neutral phases of the interstellar medium. Consequently, new dedicated GMRT \HI\ observations were carried out during November 2015; the SNR obtained is $\sqrt{3}$ better than the previous one. The KK\,69 galaxy is a good laboratory to study the processes leading from a gas-rich to a gas-poor galaxy. We aim to better understand the relationship between the star formation activity and the environmental effects over the KK\,69 galaxy evolution.\\

In the following we introduce the dwarf galaxy KK\,69 and its environment (Section~\ref{kk69}), we describe our observations and data reduction (Section~\ref{observations}), we present our results (Section~\ref{sec:results}) followed by the discussion (Section~\ref{discussion}) and the summary and outlooks (Section~\ref{summary}).

 \begin{figure*}
    \centering
     
          \includegraphics[width=0.8\linewidth]{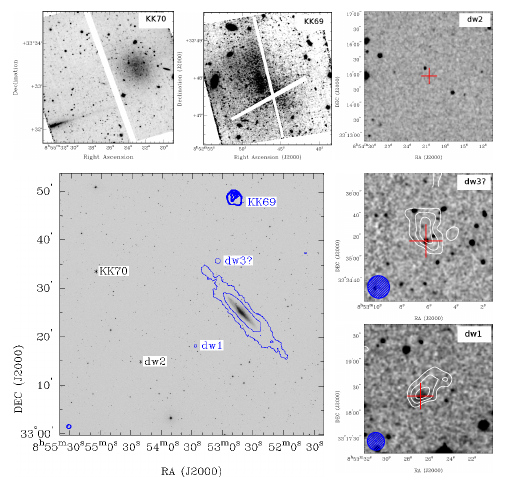}
    
   \caption{The spiral galaxy NGC\,2683 and its companion galaxies labelled. {\bf First row}, Hubble Space Telescope optical images of KK\,69 (F606W filter), KK\,70 (F814W filter) and DSS image of NGC\,2683dw2. {\bf Second row}, {\bf left:} The GMRT \HI\ intensity map of KK\,69 galaxy (thick blue contours) is overlaid onto the DSS B-band image (grey scale). The lowest contour corresponds to a columnar density of 1.5$\times 10^{19}$~atoms~$\rm cm^{-2}$, see Fig.~\ref{fig:gmrt} for more details. The VLA C+D \HI\ intensity map (thin blue contours) of NGC\,2683 is shown; the lowest contour in each VLA map corresponds to a columnar density of 4.4$\times 10^{19}$~atoms~$\rm cm^{-2}$. The \HI\ sizes of dw1 and dw3? galaxies are represented with blue circles. {\bf Second row}, {\bf right:} The VLA C+D \HI\ intensity maps of dw1 and dw3? (white contours) overlaid onto the SDSS7 R-band and DSS R-band image (grey scale) respectively. The red plus symbol marks the likely stellar component. The VLA C+D synthesised beam of 21\farcs0$\times$20\farcs0 is shown at the bottom left corner of each map.}

	\label{fig:ngc}
 \end{figure*}

\begin{table*}
  \caption{Optical properties of KK\,69 and the group member galaxies.}
    \centering
  \begin{tabular}{l@{~}c@{~}c@{~}c@{~}c@{~}c@{~}c@{~}c@{~}}
\hline
\hline
Property                            &    KK\,69       & KK\,70        & N2683       & N2683    & N2683 & NGC &    Ref        \\
                                    &                  &              & dw1        &  dw2     & dw3? & 2683&            \\
\hline
    $\alpha \rm (J2000) [h \ m \ s]$                       &   08:52:50.80   & 08:55:22.0   & 08:53:26.8   & 08:54:20.5    &08:53:06 & 08:52:40.9 & 1,2   \\
    $\delta \rm (J2000) [^o \ ' \ '']$                     &   +33:47:52.0   & +33:33:33     & +33:18:19    &+33:14:58 & +33:35:20& +33:25:02  & 1,2  \\
    $D_{\rm B26}$ [kpc]                                        &   3.7          & 1.4         & 1          &  1.2         &$-$& 37.5 & 2\\
    $a_{Holm}$ [arcmin] & 1.38 & 0.50 & 0.35 & 0.4 &$-$& 13.5 & 2\\
    $D$  [Mpc]                                         &  9.28$\pm$0.28  &  9.18$\pm$0.30& $-$ & $-$ & $-$ & 9.36$\pm$0.28 & 3   \\
    $A_{\rm B}$                                             &    0.13         &  0.13         & 0.13  & 0.12 &$-$& 0.14 & 4  \\
    $M_{\rm B}$ [mag]                                    &    -12.54       & -12.27       & -10.6  & -10.5         &$-$& -20.8 & 2   \\
   $L_{\rm B}$ [$10^7 \rm L_{\odot,B}$]                   &     1.6        &   1.2           & 0.3  & 1.7      &$-$&  3191.5&  * \\                
    $i$ [deg]                                          &      46         &  17          & 59    &  30          &$-$& 90 &2  \\
$F_{\rm H_{\alpha}}$ [$\rm erg \ cm^{-2} \ s^{-1}$]      &  (3.1$\pm$0.3)$\times 10^{-15}$   & $-$  & (7$\pm$0.9)$\times 10^{-15}$& $<$5.3$\times 10^{-16}$  &$-$& (4.5$\pm$0.8)$\times 10^{-12}$ & 5 \\ 
    $log(SFR)$ [$\rm M_{\odot} \ yr^{-1}$]                      &   $-3.8$     & $-$ & $-3.2$  & $<-4.3$    &$-$&$-0.1$&  *\\
 $v_{\rm opt}$[\kms] & $-$& $-$& 380$\pm$25 & $-$ & $-$ &335$\pm$65 & 6\\ 
\hline
  \end{tabular}
  \flushleft References: (1) \cite{2007AJ....133..715W,2010ApJS..189...37E}; (2) \cite{2013AJ....145..101K}; (3) \cite{2015ApJ...805..144K}; (4) \cite{1998ApJ...500..525S}; (5) \cite{2010AJ....140.1241K,2013AJ....146...46K,2015Ap.....58..453K,2014AstBu..69..390K,2008ApJS..178..247K}; (6) \cite{Karachentsev2015,1956AJ.....61...97H} ;(*) this paper.
\label{tab:prop}
\end{table*}




\section{KK\,69 and its environment}\label{kk69}
KK\,69, along with KK\,70, N2683dw1 and N2683dw2, are dwarf companion galaxies of the spiral galaxy NGC\,2683 \citep{2015ApJ...805..144K,2016AA...586A..98V}. All group members are shown in an optical image in Fig.~\ref{fig:ngc}. We summarise the optical properties of the member galaxies in Table~\ref{tab:prop}. \cite{2015ApJ...805..144K} determined a new distance to KK\,69 of $9.28\pm0.28$~Mpc using the tip of the red-giant branch (TRGB) method. Hereafter we adopt 9.28~Mpc as the distance to the group; at this distance 1~arcmin is equivalent to 2.7~kpc. \\




KK69 is also known as LEDA~166095 (see Fig.~\ref{fig:ngc}). It was observed with the 100\,m single-dish radio telescope at Effelsberg to obtain an integrated \HI~line flux of $F_{\rm HI}$\,=\,$2.8$~Jy\kms\ \citep{2003A&A...401..483H}. Later, GMRT observations found an \HI~flux of $3.0\pm0.3$~Jy\kms\ and a systemic velocity of $v_{\rm sys} = 462.04$~km~s$^{-1}$ \cite{2008MNRAS.386.1667B}. Based on its optical appearance, KK\,69 was first cataloged as a possible dSph galaxy by \cite{1998A&AS..127..409K}, but later on, it was proposed as a dIrr galaxy due to its \HI\ content \citep{2008MNRAS.386.1667B,2013AJ....145..101K}. The absolute B-band magnitude of $M_{\rm B}$\,=\,$-12.54$~mag means this is the brightest dwarf member of the group. The major axis diameter at the Holmberg isophote, $D_{\rm B26}$, is 3.7~kpc, and its projected distance from NGC\,2683 is 23\arcmin\ or $\sim$\,$62$~kpc. We derived a Local Group velocity, $v_{\rm LG}$\,=\,$430$~\kms, which corresponds to a Hubble distance of 5.9~Mpc adopting a value of the Hubble's constant, $H_{0}$\,$\sim$\,$73$\,\kms\,$\rm Mpc^{-1}$. The large difference between the distance determined using the TRGB and the `Hubble flow' lets us derive a KK\,69 peculiar velocity of $\sim -247$\,\kms, see also \cite{2015ApJ...805..144K}. The negative peculiar velocity implies that the motion is directed toward us.\\

KK\,70, also known as LEDA 166096 (see Fig.~\ref{fig:ngc}), is a dSph galaxy at a distance of $D_{\rm TRGB}=9.18\pm0.30$ Mpc \citep{2015ApJ...805..144K}. The absolute B-band magnitude is $-12.27$~mag, and the major axis diameter at the Holmberg isophote is 1.4~kpc, which is less than half of the value obtained for KK\,69. The projected distance to NGC\,2683, is 30\arcmin\ or $\sim$82~kpc. An \HI~non-detection in KK 70 led to an estimation of the upper limit on the \HI~flux as $F_{\rm HI}=0.48$~Jy~\kms. This translates into an \HI~mass limit of $M_{\rm HI}=10^{7} \rm M_{\odot}$ at the distance of the galaxy \citep{2014AJ....147...13K}.\\


N2683dw1 and N2683dw2, hereafter dw1 and dw2, are dIrr and dSph galaxies, assumed to be galaxy members of the NGC\,2683 galaxy group \citep{Karachentsev2015}. Their $M_{\rm B}$ and $D_{\rm B26}$ imply that they are the faintest and smallest members in the group, see Table ~\ref{tab:prop}. The dw1 galaxy is at a projected distance of $\sim$7\arcmin\ or $\sim$\,$19$\,kpc while dw2 is at a projected distance of $\sim$10\arcmin\ or $\sim$\,$27$\,kpc from the spiral galaxy. The galaxy dw1 is detected in \HI; the results are shown in Section~\ref{sec:results}. Considering the r.m.s. of 1~mJy~$\rm beam^{-1}$ per 5~\kms\ channel width in \cite{2016AA...586A..98V}, we estimated an \HI\ flux upper limit of $F_{\rm HI}=0.09$~Jy~\kms\ for dw2 assuming a spectral width of  $\sim$30~\kms.\\

NGC\,2683 is an edge-on spiral galaxy at a distance of $9.36 \pm 0.28$~Mpc. It is the most massive galaxy in this group. The $F_{\rm HI}$ is 101.4\,Jy\kms, the linewidths at 20$\%$ and 50$\%$ of the flux density are $w_{20}$\,=\,$450$\kms\ and $w_{50}$\,=\,$426$\kms \citep{2016AA...586A..98V}.
 Using the group distance, the calculated value of its total \HI\ mass is found to be $M_{\rm HI}=2.3 \times 10^9\, \rm M_{\odot}$. We derived a Local Group velocity of $v_{\rm LG}$\,=\,$376$\kms, which corresponds to a Hubble distance of $\sim$5\,Mpc. The large difference between the distance determined using the TRGB and the `Hubble flow' let us derive a NGC\,2683 peculiar velocity of $-340 \, \rm km \, s^{-1}$, see also \cite{2015ApJ...805..144K}. \cite{2010AJ....140.1241K} have estimated the $\rm H_{\alpha}$ flux of $F_{\rm H_{\alpha}}$\,=\,4.5$\times$\,$10^{-12}$~$\rm erg\,cm^{-2}\,s^{-1}$ and the star formation rate of $SFR=10^{-0.1} \rm M_{\odot} \ yr^{-1}$.\\
 
We estimated, for each galaxy, the luminosity in the blue band using the expression:
\begin{displaymath}
L_{\rm B}=D^2\times10^{10-0.4(m_{\rm B}-A_{\rm B}-M_{\rm B,\odot})},
\end{displaymath}  
where $D$ is the distance, $m_{\rm B}$ is the apparent magnitude, $A_{\rm B}$ the galactic extinction and $ M_{\rm B,\odot}$ is the absolute magnitude of the Sun. We have also determined the SFR, when was possible, using the relation $SFR[\rm M_{\odot}\,yr^{-1}]$~$= 0.945\times 10^9\,F_{\rm H\alpha}\,D^2$ \citep{1998ARA&A..36..189K}.

 \section{Observations and data Reduction}\label{observations}
 
\begin{table*} %
  \centering
  \caption{GMRT observing parameters.}
  \begin{tabular}{lccc}
    \hline
    \hline

    Date of observations               & 14-15 Nov 2015 & 16-17 Nov 2015 & 2 Jan 2005\\
    Time on source [min]                 & 322            & 360            & 258\\
    Pointing centre &\\
    \,\,\ $\alpha \rm (J2000) [h \ m \ s]$       &\multicolumn{2}{c}{8:52:50.7}& 8:52:50.69 \\
    \,\,\ $\delta \rm (J2000) [^o \ ' \ '']$     &\multicolumn{2}{c}{+33:47:51.9
    }& +33:47:52\\
    Flux of calibrator 3C147 [Jy]    & 22.2$\pm$0.3& 22.4$\pm$0.3 &22.1$\pm$0.02\\
    Flux of calibrator 3C286 [Jy]    & 15$\pm$0.2 & 15.2$\pm$0.2  &14.7$\pm$0.03\\
    Flux of calibrator 741+312  [Jy] & 2.1$\pm$0.03 & 2.3$\pm$0.03&2.2$\pm$0.01\\
    Centre frequency [MHz]             & \multicolumn{2}{c}{1418.3524} & 1417.1983\\
    Bandwidth [MHz]                    & \multicolumn{2}{c}{4.166} & 1\\
    No. of channels                    & \multicolumn{2}{c}{512} & 128\\
    Channel width [MHz]                &\multicolumn{2}{c}{0.008138} & 0.007812\\
    Velocity resolution [\kkms]        & \multicolumn{2}{c}{1.7} & 1.65 \\
    \hline
    \end{tabular}
\label{tab:GMRTobservations}
\end{table*}  
    
\begin{figure}
\centering
\includegraphics[width=0.45\textwidth]{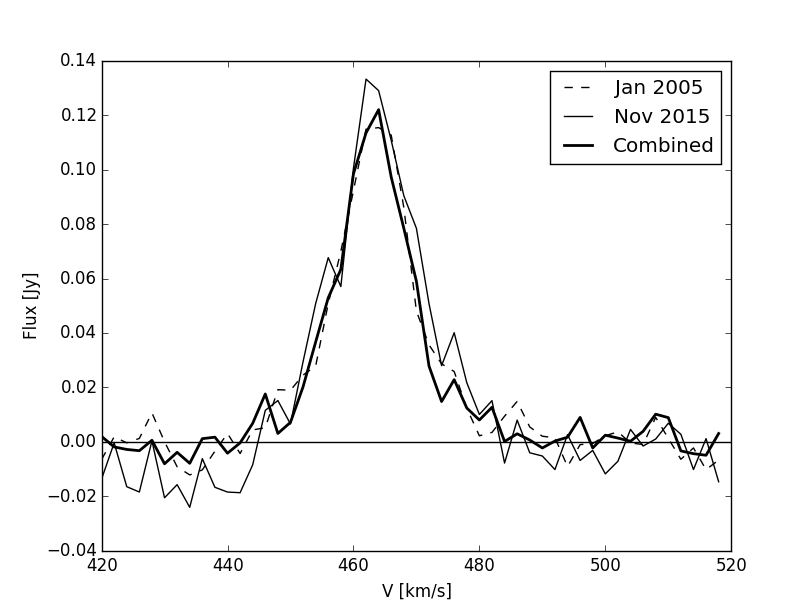}\\
\caption{The integrated \HI\ spectrum of the KK\,69 galaxy using GMRT observations in 2005 (dashed black line), 2015 (black line), and the combination of both (thick black line). The velocity range over which we have detected the \HI\ emission in KK 69 is $\sim 440$~to~490\kms.}
\label{fig:spectra}
\end{figure}


We observed KK\,69 with the Giant Metrewave Radio Telescope (GMRT) in the L band for a total time of 11.2~hours. The observations were carried out during November 2015 in the spectral zoom mode with the GMRT-Software backend. We used a total $\sim$4~MHz bandwidth with 512 channels, which results in a spectral resolution of $\sim$8.13~kHz, equivalent to 1.7\kms\ per channel. The flux calibrators 3C286 and 3C147 were observed at the beginning and the end of the run for $\sim$15 min each. The source 0741+312 was used as the phase calibrator and was observed for $\sim$5 min between the target scans of $\sim$45 min on KK\,69; see Table~\ref{tab:GMRTobservations} for details.\\

Previous data, as well as the new ones, were flagged and calibrated using the ``Flagging and calibration pipeline for GMRT data'' \citep[FLAGCAL]{2011ascl.soft12007P} and combined during the imaging process using {\sc miriad} software package \citep{1995ASPC...77..433S}. In addition, we extensively used the Astronomical Image Processing Software (AIPS) for further analysis and the multi-wavelength imaging software {\sc kvis}, part of the {\sc karma} package \citep{1996ASPC..101...80G} for visualisation. GMRT does not do online Doppler tracking. Thus, the AIPS task CVEL was implemented to apply the Doppler shift corrections. We used the {\sc miriad} task {\sc uvlin} to subtract the continuum.\\
To probe the \HI\ structure at various spatial scales, we imaged the visibility cube at different spatial resolutions by appropriately setting the tapering limits. However, since the galaxy is very faint, for all practical purposes of the current study, we used a low spatial resolution image cube. The \HI\ map synthesised beam for natural weighting, considering up to 5~k$\lambda$ baselines, is 50\farcs39$\times$49\farcs57 ($2.3\times 2.3$~kpc for a distance of 9.28~Mpc). The r.m.s. in the final combined data cube is $\sim$1~mJy~beam$^{-1}$ per 2\kms\ channel in the center of the field, almost two times better than previous observations \citep{2008MNRAS.386.1667B}. The 3$\sigma$ $N_{\rm HI}$ and \MHI\ limits over ten channels are 1.5$\times$10$^{19}$~atoms~cm$^{-2}$ and $\sim 10^{7}$\Msun (assuming the gas fills the beam).\\


We have also made use of the \HI\ VLA reduced data cubes published by \cite{2016AA...586A..98V}. The authors studied in detailed the galaxy NGC\,2683 and left aside the dwarfs galaxies close to it, which some of them are present in the C+D final cube. The final C+D cube mentioned in their work results as the combination of both VLA array observations. The r.m.s. noise level is 0.3~mJy~$\rm beam^{-1}$ in a 10.3\kms\ channel with a resolution of 21\farcs0$\times$20\farcs0.

\section{Results}\label{sec:results}

The global \HI\ spectrum of KK\,69 has a narrow Gaussian shape, and the \HI\ emission is detected from $\sim$440 to 490\kms, see Fig.~\ref{fig:spectra}. After the primary beam correction of our image cube, we measured a total \HI\ flux of $F_{\rm HI}$=2.1$\pm$0.3~Jy\kms; this value corresponds to a total \HI\ mass of $M_{\rm HI}$=4.2$\times 10^7$M$_{\odot}$, at a distance of 9.28~Mpc. The line widths at 20$\%$ and 50$\%$ of the peak flux density are $w_{20}$=22\kms\ and $ w_{50}$=12\kms\, with a systemic velocity of $\sim$464\kms. These estimations were done using the {\sc mbspect} task in {\sc miriad}. The \HI\ properties of KK\,69 derived from the GMRT observations are listed in Table~\ref{tab:HI}.\\

\begin{figure*}
\centering
\includegraphics[width=0.8\textwidth]{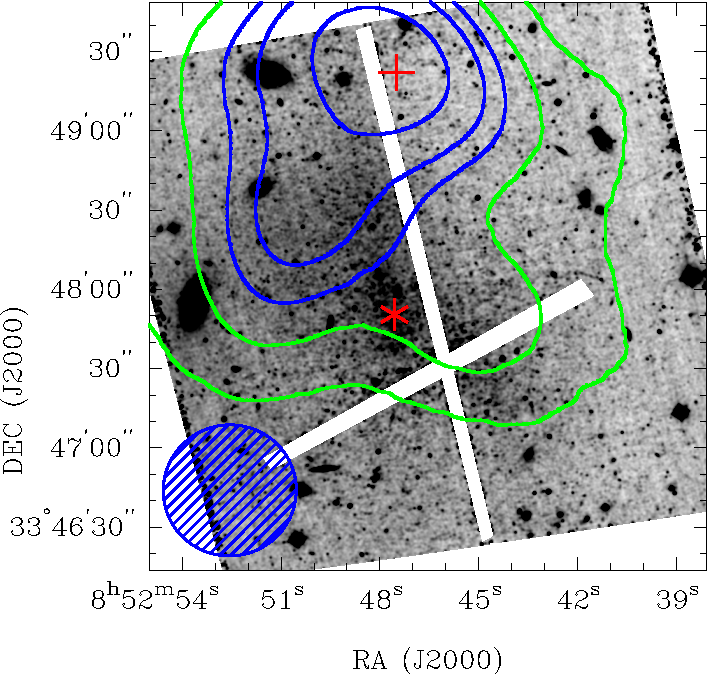}
\caption{The dwarf galaxy KK\,69. The GMRT \HI\ intensity map of KK\,69 is overlaid onto the Hubble Space Telescope image F606W band (grey scale). The \HI\ contour levels are (1, 4, 6, 9, 12 )$\times$~30~mJy~$\rm beam^{-1}$\kms\ or 1.5$\times 10^{19}$~atoms~$\rm cm^{-2}$. The center of the stellar distribution (star symbol) and the peak of the \HI\ gas (plus symbol) are not coincident; the offset between the two is $\sim$ 1.6 arcmin or $\sim$ 4.3 kpc for a distance of $D=9.28$ Mpc. The synthesised beam (50\farcs39$\times$49\farcs57) is shown in the bottom left corner.} 
\label{fig:gmrt}
\end{figure*}

\begin{figure*}
 \centering
      \includegraphics[scale=0.25]{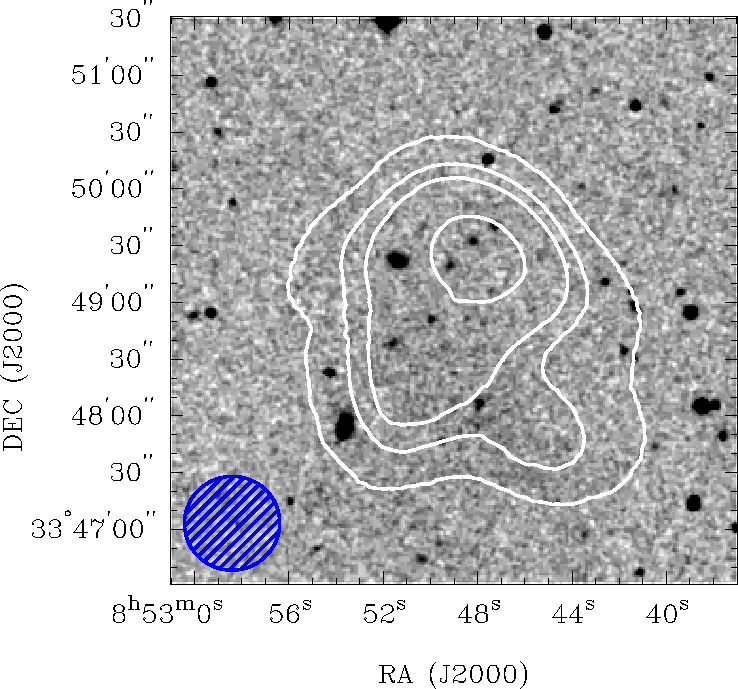}
  \includegraphics[scale=0.3]{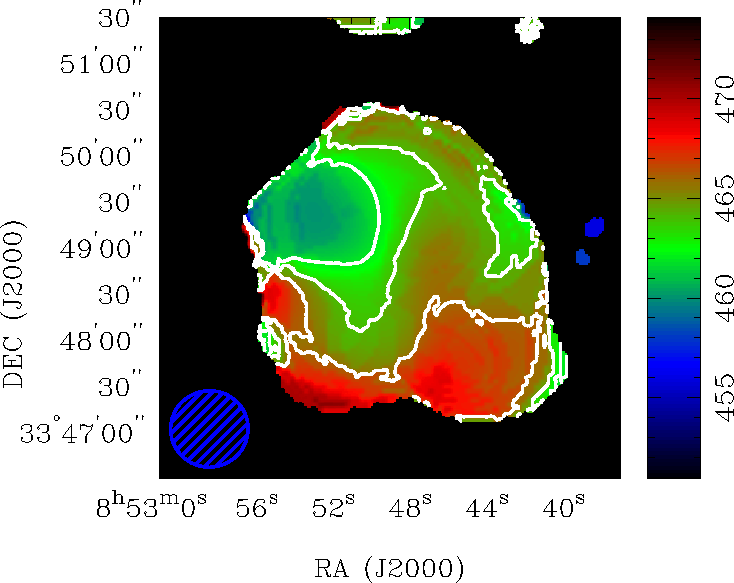}
   \includegraphics[scale=0.3]{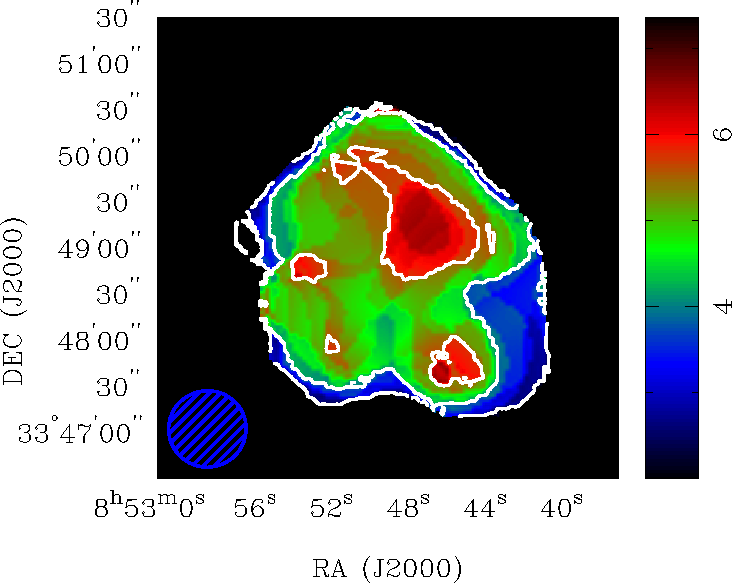}

\caption{GMRT \HI\ moment maps of the dwarf galaxy KK\,69 in the velocity range from 440 to 490\kms. {\bf First panel:} The \HI\ intensity map; the contour levels are the same as in Fig.~\ref{fig:gmrt}. {\bf Second panel:} Mean \HI\ velocity field; the contour levels are 462, 464, 466\kms . {\bf Third panel:} Mean \HI\ velocity dispersion; the contour levels are 2.3, 4, 5.6\kms . The synthesised beam (50\farcs39$\times$49\farcs57) is shown in the bottom left corner.}
\label{fig:mom0}
\end{figure*}

 In order to study the \HI\ gas distribution and kinematics in KK\,69, we derived different moment maps. Before we did that, we applied a three-point spectral Hanning smoothing to the original spectral cube. We made use of the moment routine in the {\sc miriad} software with a threshold clip of 3 times the r.m.s of the smoothed cube, see Figs.~\ref{fig:gmrt}~\&~\ref{fig:mom0}. The \HI\ distribution in KK\,69 is not symmetrical. The moment zero map reveals a shift between the center of the \HI\ distribution and the peak of the \HI. Moreover, the KK\,69 \HI\ gas is offset towards the northwest of the stellar body, and the \HI\ contours are more compressed, see Figs.~\ref{fig:gmrt}, while in the south the \HI\ emission is more extended and diffuse. The position of the \HI\ peak is $\alpha,\delta \rm (J2000) =  08^{h} 52^{m} 47.49^{s}, +33^{o} 49^{'} 21\farcs98$. The displacement measured between the stellar high-density and the \HI\ high-density regions is $\rm \sim$1.6\arcmin or $\sim$4~kpc, assuming the group distance of 9.28~Mpc. The velocity field presented in Fig.~\ref{fig:mom0} shows evidence of a weak velocity gradient, from $\sim$450~to~470\kms, across the extent of KK\,69 with a mean velocity dispersion around of 5.5\kms. If we assume that the kinematics of this galaxy is rotationally supported, we can derive a total dynamical mass (M$_{\rm dyn}$\footnote{$M_{\rm dyn}$=2.31$\times$10$^5$~r[kpc]~($v_{\rm rot}/\rm sin(i_{\rm HI}))^2$.}) of 6$\times$10$^{7}$\Msun. This is just $\sim$1.5 larger than the total \HI\ mass of KK\,69, which means that the galaxy is not in dynamical equilibrium. Probably, the combination of rotation and turbulence motions describes the velocity field of KK\,69 in Fig.~\ref{fig:mom0}. High-velocity resolution and high-angular resolution are needed to understand the kinematics of this galaxy better.\\ 

After the inspection of the VLA C+D image cube, we discovered the \HI\ counterpart of dw1, and a high-velocity cloud/tiny new galaxy (hereafter dw3?). dw3? is located, in projection, near the north-eastern end of NGC\,2683. We used the primary beam corrected image cube to measure the \HI\ flux in each galaxy. The \HI\ flux and the systemic velocity of dw1 are $\sim$0.3~Jy\kms\ and $v_{\rm sys}=421$\kms. The total \HI\ mass is $M_{\rm HI}=0.6\times 10^7 \rm M_{\odot}$.  The \HI\ intensity distribution map overlaid onto the DSS R-band magnitude image shows that the stellar and the \HI\ high-density region are coincident, but the centers of the \HI\ distribution and the optical counterpart are offset. A tail feature is observed towards the north-western part of the galaxy, see Fig.~\ref{fig:ngc}. We derived a Local Group velocity $v_{\rm LG}=374$~\kms, which corresponds to a Hubble distance of 5~Mpc, based on Hubble's constant. The large difference between the distance, considering dw1 as a group member, and the `Hubble flow' let us derive a peculiar velocity of $-345$~\kms. The total \HI\ flux measured for dw3? is $\sim$0.6~Jy\kms. Assuming the group distance, the total \HI\ mass is $M_{\rm HI}=1.2\times 10^7 \rm M_{\odot}$. Due to the better quality of the optical image given by the R-band of SDSS-DR7 in comparison with the DSS one, we overlaid this onto the \HI\ intensity distribution map of dw3?; the optical image shows a compact reddish/blueish optical counterparts which could be the tip of a more extended and diffuse stellar component. We marked this with a red cross in Fig.~\ref{fig:ngc}. The \HI\ properties derived from the VLA C+D image cubes are listed in Table~\ref{tab:HI}.

\begin{table*}
  \caption{\HI\ properties of the KK\,69 and the group member galaxies detected, assuming a group distance of 9.28~Mpc.}
  \centering
  \begin{tabular}{l@{~~}c@{~~}c@{~~}c@{~~}c@{~~}}
\hline
\hline 
                    Property                       &  KK\,69 & dw3? & dw1 & NGC\,2683\\
       
\hline
    \HI\ peak                                                          &&&&\\
  \, $\alpha \rm (J2000) [h \ m \ s]$              & 08:52:47.49   & 08:53:06     & 08:53:26     & 08:52:40.9\\
 \,  $\delta \rm (J2000) [^o \ ' \ '']$            & +33:49:21.98  & +33:35:20    & +33:18:25    & +33:25:02\\
   $F_{\rm HI}$ [Jy \kms]                              & 2.1$\pm$0.3   & $\sim$0.6 & $\sim$0.3   &$\sim$104 \\
   $M_{\rm HI}$ [$10^7 \rm M_{\odot}$]                 & 4.2$\pm$0.6   & $\sim$1 & $\sim$0.6   &$\sim$210\\
   $M_{\rm HI}/L_{\rm B}$ $[\rm M_{\odot}]$            & 2             & $-$          &  $-$         & 0.05\\
   $D_{\rm HI}$ $\rm [arcmin]$                         & 2$\times$1.40 & 0.9$\times$0.6 & 0.9          &26.5$\times 5$\\
  $D_{\rm HI}$ $\rm [kpc]$                             &5.4$\times$3.8 & 2.4$\times$1.6  & 2.4     &72$\times$13\\
   \vsys [\kms]                                        &$\sim$464  &$\sim$467     & $\sim$421    & $411$\\
   $w_{50}$ [$\rm km \ s^{-1}$]                        & 12            & 23          &  15         & $426$     \\
    $w_{20}$ [$\rm km \ s^{-1}$]                       & 22            & 35           & 31           & $450$  \\
  $v_{\rm LG}$ [$\rm km \ s^{-1}$]                     &430            & 422       &   374        &376\\
   $v_{\rm pec}$ [$\rm km \ s^{-1}$]                   & $-247$        & $-251$       & $-341$       &$-340$     \\
\hline  
  \end{tabular}
\label{tab:HI}
\end{table*}

\subsection{The state of the HI gas in KK\,69}

Theoretical understanding suggests that the interstellar gas would settle in one of the two stable phases, i.e., Cold Neutral Medium (CNM) or Warm Neutral Medium (WNM) under local thermal equilibrium \citep{field69b,mckee77,wolfire95a,wolfire03}. The CNM would have a temperature of $\leq$1000~K with high particle density ($\sim$1-100~cm$^{-3}$), whereas the WNM would have a much higher temperature $\geq$5000~K with low particle density ($\sim$0.1-1~cm$^{-3}$), see \cite{dickey78,payne83,heiles03a,roy06}. Any gas with an intermediate temperature would quickly move to one of these stable phases by thermal runaway processes. Different emission studies have been carried out in order to identify the existence of the two phases of the ISM.
 \citet{young96} decomposed the line-of-sight \HI\ emission spectra of many dwarf galaxies into multiple Gaussian components and found clear shreds of evidence of two components, one with a broad width ($\sim$8-12\kms) and other with a relatively narrow width ($\leq$6\kms). Even though the widths correspond to kinetic temperatures higher than the ones expected from theoretical models, the authors attributed this to a non-thermal broadening by turbulence. The decomposition of the spectra could be done considering either the spatially resolved line profiles or stacking the spectra to create a super-profile.\\
 
\begin{figure*}
\includegraphics[width=1.0\textwidth]{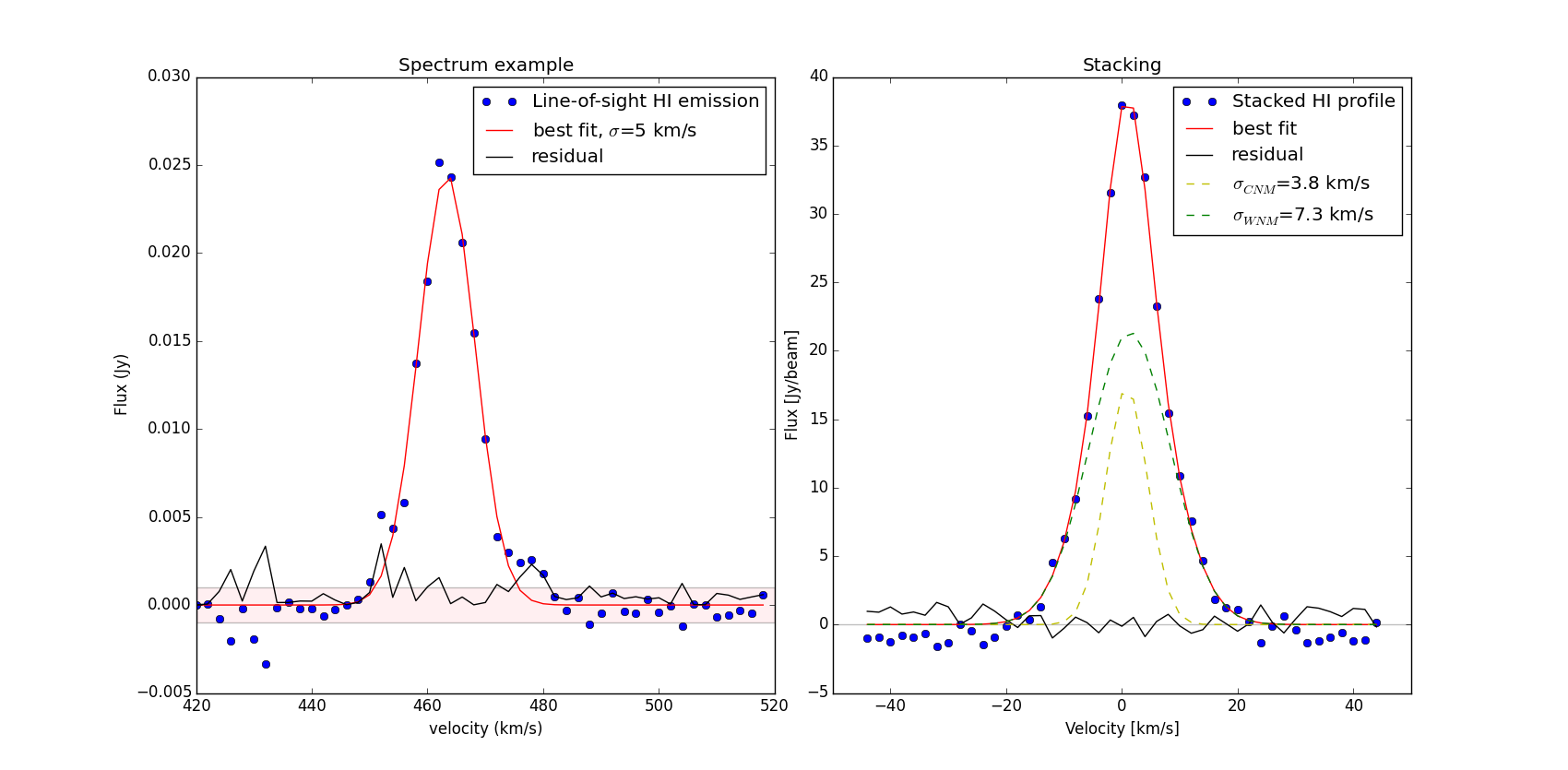}
\caption{{\bf Left panel}: An example \HI\ spectrum from the spectral cube of KK\,69. This spectrum has an SNR of 25 and is clearly well fitted by a single Gaussian profile with more than 95\% confidence level in the F-test. The pink coloured area marks the $\pm$3$\sigma$ region. {\bf Right panel}: The stacked \HI\ spectrum of KK\,69. The blue solid points represent the stacked spectrum and the red solid line indicates a double-Gaussian fit to the spectrum. The yellow dashed line represents the narrow Gaussian component (CNM) while the green dashed line represents the wide Gaussian component (WNM). The fit residuals are shown with a black line.}
\label{fig:fit-comp}
\end{figure*}

We first implemented the {\tt multigauss} routine from \cite{2016MNRAS.456.2467P}, adopting the same approach as in previous studies. To guarantee reliable fits, we have only decomposed the spectra with an SNR$>$10, and to avoid any false positive, we use a 95\% confidence level for F-test. The high SNR requirement severely restricts the total area over which we could apply the Gaussian decomposition method; in fact, it was equivalent to $\sim$4 synthesised beam area. The CNM component was found within the blue contours, as shown in Fig.~\ref{fig:gmrt}. A single Gaussian component best fits most of the KK\,69 spectra with a $\sigma_{\rm HI} \leq$6\kms, see Fig.~\ref{fig:fit-comp}, with mean velocity dispersion of 5.5\kms. A very few line-of-sights were best fitted by Gaussian profiles of widths 8\kms$< \sigma_{\rm HI} <$12\kms. This implies that a better signal-to-noise ratio, as well as a better angular resolution, are needed to identify the WNM component directly.\\

Hence, to get a better signal-to-noise, we stacked the line-of-sight \HI~spectra. We adopted a similar approach as it is described in \cite{2013ApJ...773...88S}. We fitted the line-of-sight \HI~spectrum with a Gaussian Hermite Polynomial to locate the centroid of individual spectra. We used the information of the centroids to align all the spectra to a common velocity and stacked them together to produce a high SNR \HI~spectrum. The thinness of the resultant stacked spectra, in comparison with the total \HI\ profile, implies that no artificial broadening is being created by asymmetric wings. In Fig.~\ref{fig:fit-comp}, we plot the resulting stacked spectra for KK\,69.  
To produce a meaningful fit of the spectra by Gaussian Hermite Polynomial, a  minimum SNR of 5 is required   \citep{deblok08}. The decomposition of the stacked spectrum was carried out using both a single and a double Gaussian. According to the F-test, the two-component Gaussian fit is preferred to a single Gaussian component with a confidence level of better than 95$\%$. In Fig.~\ref{fig:fit-comp} we plot the resulting decomposed components, the narrow and the broad components with velocity dispersion of $\sigma_{\rm N}$=$3.8 \pm 0.3$\kms\ (T$\sim$840 K\footnote{(3/2)kT=(1/2)$m_p \sigma^2$.}),  and $\sigma_{\rm B}$=$7.3 \pm 0.3$\kms\ (T$\sim6300$ K), respectively. As can be seen from Fig.~\ref{fig:fit-comp}, the residuals look like noise without any significant feature indicating a double Gaussian to be a good representation of the stacked spectra.  We found that $\sim 30\%$ of the total \HI\ gas is well described by the narrower component while $\sim 70\%$ by the broader component.

\section{Discussion}\label{discussion}

We found evidence that KK\,69 is a dwarf transition galaxy being transformed from a gas-rich to a gas-poor galaxy. In the following paragraphs we discuss our results taking into account the relation between the interstellar medium and the star formation activity in the galaxy, the role played by the environment and the influence of these phenomena in the galaxy's evolution.\\

To characterise in detail the \HI\ gas of KK\,69, we run a multi-Gaussian decomposition code and a stacking routine. From the direct decomposition method, we found that most of the \HI\ line-of-sight spectra are best described by a single Gaussian with a velocity dispersion around 5.5\kms, which probably corresponds to the CNM component. Certainly, this value is high in comparison with the values present in the literature ($\sim$3-4\kms), but it is similar to the ones obtained for the galaxies GR8, M\,81\,Dwarf\,A and NGC\,4190 which are the faintest and smallest dwarf galaxies in \cite{2012ApJ...757...84W} sample. The stacking method revealed the presence of two components, one described by a narrow profile and another one by a broad profile, with velocity dispersion of $\sigma_{\rm N}=3.8 \pm 0.3$\kms\ and $\sigma_{\rm B}=7.3 \pm 0.3$\kms, respectively. The $\sigma_{\rm N}$ value is in agreement with the typical velocity dispersion of the cold phase ($\sim$4.5\kms); see \cite{2012ApJ...757...84W}. The $\sigma_{\rm B}$ is smaller than the typical value found for the warm phase by the same authors, but it is similar to the velocity dispersion detected for the WNM component in the dwarf transition galaxy Leo T \citep{2008MNRAS.384..535R,2018A&A...612A..26A}. The main issue regarding the WNM detection is the fact that this component is extended and diffuse.
In Fig.~\ref{fig:spectra}, we compared the GMRT integrated \HI\ spectrum of the KK\,69 galaxy obtained from 2005 and 2015 observations as well as the combinations of both of them. Even though we carried out a similar data reduction procedure, the flux we measured is 2.1$\pm$0.3~Jy\kms\, which is somewhat lower than the previous value of 3.0$\pm$0.3~Jy\kms\ published by \cite{2008MNRAS.386.1667B}. The total \HI\ flux recovered
after combining the GMRT observations is lower than the detected \HI\ flux using the single-dish \citep{2003A&A...401..483H}. We are probably facing the fact that due to GMRT’s lack of short baselines, the GMRT observations struggle to detect the warmer component of the \HI\ gas, see also \cite{2008MNRAS.384..535R}. 
It is clear that the decomposition of the \HI\ line-of-sight spectra into the CNM and the WNM is not straight forward for this galaxy, but we found evidence that a huge amount of \HI\ gas lies in the CNM phase.\\

The CNM component is concentrated beyond the bulk of the stellar population, see Fig.~\ref{fig:gmrt} blue contours. The CNM distribution in the dwarf galaxies present in \cite{2012ApJ...757...84W} is quite different from the CNM distribution in KK\,69. While in the first, the cold component is typically located at \HI\ column densities above 10$^{21}$~atoms~cm$^{-2}$, in KK\,69 it is found at \HI\ column densities above 10$^{20}$~atoms~cm$^{-2}$; a shared characteristic with the galaxy Leo T \citep{2018A&A...612A..26A}. The \HI\ high-density region of KK\,69 contains a CNM column density\footnote{$N_{\rm gas}= 1.33 \times N_{\rm HI} \approx 5\times 10^{21}$  $\rm atoms \ cm^{-2}$ corrected for He abundance.} level of 2.4$\times$10$^{20}$~atoms~cm$^{-2}$. This value is well below the canonical threshold of star formation of 10$^{21}$~atoms~cm$^{-2}$, at a linear resolution of 500~pc \citep{1987NASCP2466..263S}. Probably, since the observed column density is an average over the telescope beam (2.3$\times$2.3~kpc) the local column density is higher than the obtained value.\\
Observation carried out with BTA 6-m telescope by \cite{2013AJ....146...46K}, detected H$_{\alpha}$ emission coming from a point source plus a diffuse component, located at the north-west of the stellar high-density region. For KK\,69, the authors derived a star formation rate of 10$^{-3.8}$M$_{\sun}$~yr$^{-1}$. We looked into the GALEX images, and we found a hint of near-UV (NUV) and far-UV (FUV) emission nearby the \HI\ high-density region.
Considering the SFR we could estimate the gas depletion timescale\footnote{$\tau_{\rm gas}= 1.32 \times M_{\rm HI} /SFR$; the factor 1.32 takes into account the presence of He.} of $\tau_{gas}$=86.4~Gyr. If the SFR remains constant, the galaxy will need six times the age of the universe to transform all the \HI\ gas into stars. This value is not as high as the one found in Pegasus ($\tau_{\rm gas}$=3220~Gyr) in the Local Group, UGCA 365 ($\tau_{\rm gas}$=1250~Gyr) and UKS 1424-469 ($\tau_{\rm gas}$=534~Gyr) in the Centaurus A group \citep{2009ApJ...696..385G}, but is still high in comparison with the $\tau_{\rm gas}$ estimate for galaxies forming stars more actively. The second method we implemented was the stacking process. From this method, we found that $\sim$30\% of the total \HI\ gas in KK\,69 remains in the CNM component. Although the KK\,69 galaxy has a substantial reservoir of \HI\ gas, in particular CNM which is the main fuel for star formation, a sluggish star formation activity seems to characterise this galaxy.\\
The estimated metallicity for KK\,69 is $12+log(O/H)=7.5$, following the relation proposed by \cite{2010MNRAS.403..295E}. This low metalled means neither the time-averaged star formation rate chemical evolution has been lower than usual nor the galaxy has lost metal-enriched gas through galactic winds/supernovas. The comparison with the size of the \HI\ gas offset from the stellar body observed in the galaxy Phoenix \citep{2007ApJ...659..331Y} as well as the discussed properties of star formation activity in this galaxy, suggests that this latter scenario is less plausible.\\

The low star formation activity seems to be a general feature in the other dwarf galaxies member of the group.
The dwarf galaxies KK\,70, dw2, and dw1 were also observed with the BTA 6-m telescope. No signs of H$_{\alpha}$ emission were detected for KK\,70 by \cite{2013AJ....146...46K}, and a H$_{\alpha}$ upper limit flux was estimated for dw2 \citep{2019AstBu..74....1K}. On the contrary, H$_{\alpha}$ emission was detected in dw1 \citep{2014AstBu..69..390K}, and this is in agreement with a clear GALEX counterpart. We did not find evidence of UV emission in dw3? after the inspection of GALEX images.\\

KK\,69 is evolving in a particular scenario. The galaxy KK\,69 belongs to a small group of galaxies located in the front edge of the Gemini-Leo Void. The main galaxy in this group is NGC\,2683, a spiral edge-on galaxy, with some resemblances to the Milky Way and a bunch of neighboring dwarf galaxies. Considering the velocity and its projected distance, dw1 is the closest galaxy to NGC\,2683. In Fig.~\ref{fig:ngc}, we show the \HI\ distribution of the galaxy dw1. The stellar body and the \HI\ high-density region are coincident, but the \HI\ gas distribution shows a tail appearance towards the north-western direction of the galaxy. The subsequent closest galaxies are dw2, with no \HI\ detection, and dw3?. What is the nature of dw3? ? Is it a high-velocity cloud, or is it a dwarf galaxy close to NGC\,2683?. The \HI\ velocity of dw3? is in agreement with the \HI\ systemic velocity of KK\,69. The \HI\ size and the \HI\ mass of dw3? are bigger than the values obtained for dw1. The SDSS7 R-band image shown in Fig.~\ref{fig:ngc} exhibits, within the \HI\ distribution, a compact optical counterpart, which could be thought of as the tip of a more extended and diffuse stellar component. From the comparison with the other dwarf galaxies members of the group, we proposed that dw3? is a new dwarf galaxy. Deep optical images are needed to confirm or dismiss the idea of dw3? being a dwarf galaxy.\\
Assuming that the virial theorem applies in this system, we roughly estimate a virial radius of 240 kpc \citep{2001A&A...367...27W,2007gaun.book.....S}. Thus, KK\,69, as well as the other dwarf galaxies companions of NGC\,2683, are well within the virial radius of NGC\,2683. Probably, the interaction of dw1 and dw3? with NGC\,2683 is the main mechanism leading the observed bent in the southwestern and northeastern part of the \HI\ disk of NGC\,2683. In KK\,69, the \HI\ gas is not symmetrically distributed. It is concentrated towards the north-west where the outermost \HI\ contours are more compressed than elsewhere in the galaxy, and in the south-east part of the galaxy, the \HI~emission is more extended and diffuse, see Fig.~\ref{fig:mom0}. The observed morphological segregation in groups such as the Local Group, Centaurus, and Sculptor Group implies that the \HI\ gas is removed from the dwarf galaxies that interact closely with massive galaxies. Given the fact that KK\,69 is a gas-rich galaxy well within the virial radius of NGC\,2683, the offset observed as well as the compressed contours are likely produced by the interaction of the \HI\ gas disk with the intragroup medium. The idea of KK\,69 infalling onto the group is the most probable.\\


\section{Summary and outlooks}\label{summary}

We carried out this study using new dedicated \HI\ observations of the dwarf galaxy KK\,69, obtained with the GMRT.
By means of a multi-Gaussian decomposition code and a stacking routine, we have found that a considerable amount of \HI\ gas remains in the CNM phase ($\sim$30\%). Even though the galaxy seems to have a huge reservoir of the considered main fuel for star formation, it is not forming stars actively. The low star formation activity is a common characteristic shared with the other dwarf galaxies members of the group.\\

 The GMRT \HI\ observations revealed an offset between the \HI\ gas distribution and the stellar body. Assuming a group distance of 9.28 Mpc, the size of the offset is found to be $\sim$4~kpc. The measured \HI\ flux is 2.1~Jy~beam$^{-1}$\kms\ and thus the total \HI\ mass is $M_{\rm HI}=4.2\times 10^{7} \rm \, M_{\odot}$. This established that KK\,69 is the most \HI\ massive galaxy in this group after NGC\,2683. The velocity field presents a weak velocity gradient, but the mean dispersion velocity of the \HI\ gas has a similar amplitude. The galaxy is not under dynamical equilibrium. High-angular resolution observations are needed to understand the kinematics of KK\,69 better.\\
 
The nearest large neighbour of KK\,69 is the spiral galaxy NGC\,2683 ($M_{\rm HI}=2\times 10^{9} \rm \,M_{\odot}$) with signs of recent external gas accretion. After the inspection of the VLA C+D image cube published by \cite{2016AA...586A..98V}, we found the \HI\ counterpart of a dwarf galaxy member of the group (dw1) and a possible new dwarf galaxy (dw3?) located at the north-east of the NGC\,2683 \HI\ disk. In both objects, a displacement is observed between the centers of the \HI\ gas and the stellar component. Moreover, dw1 shows a kind of a tail feature in the north-western part. All the dwarf galaxies mentioned are well within the virial radius of NGC\,2683. We proposed that tidal interactions are the main responsible mechanism of the observed offset between the stellar component and the \HI\ distribution in the dwarf galaxies member of this group. KK\,69 is likely a dwarf transition galaxy in which the \HI\ gas is being stripped from the main body.\\
 
 Upcoming surveys as for example, WALLABY \citep{2012PASA...29..359K}, are expected to discover a large number of transition galaxies and with it the possibility of characterising their evolutionary states. X-ray observations are also needed in order to obtain information about the existence of intragroup hot gas and thus the possibility to study the importance of ram-pressure stripping process in low-density environments.

\section*{Acknowledgements}
The radio data presented here were obtained with the Giant Metrewave Radio Telescope (GMRT). The GMRT is operated by the National Centre for Radio Astrophysics of the Tata Institute of Fundamental Research.
JS is greateful to the GMRT and NCRA staff for their hospitality during her stay. This research has made use of the NASA/IPAC extragalactic database (NED) that is operated by the Jet Propulsion Laboratory, California Institute of Technology,
under contract with the National Aeronautics and Space Administration and of SIMBAD data base, operated at CDS, Strasbourg, France. This work was partially supported by FCAG-UNLP
and ANPCyT project PICT 2012-878.

\bibliographystyle{spr-mp-nameyear-cnd}
\bibliography{biblio-kk69.bib}  

\end{document}